# Supporting Bandwidth Guarantee and Mobility for Real-Time Applications on Wireless LANs

Srikant Sharma Kartik Gopalan Ningning Zhu
Gang Peng Pradipta De Tzi-cker Chiueh
{srikant,kartik,nzhu,gpeng,prade,chiueh}@cs.sunysb.edu
Computer Science Department
Stony Brook University
Stony Brook, NY - 11794

## Abstract

*The proliferation of IEEE 802.11-based wireless LANs opens up avenues for creation of several tetherless and mobility oriented services. Most of these services, like voice over WLAN, media streaming etc., generate delay and bandwidth sensitive traffic. These traffic flows require undisrupted network connectivity with some QoS guarantees. Unfortunately, there is no adequate support built into these wireless LANs towards QoS provisioning. Further, the network layer handoff latency incurred by mobile nodes in these wireless LANs is too high for real-time applications to function properly. In this paper, we describe a QoS mechanism, called* Rether, * *to effectively support bandwidth guarantee on wireless LANs. Rether is designed to support the current wireless LAN technologies like 802.11b and 802.11a with a specific capability of being tailored for QoS oriented technology like 802.11e. We also describe a low-latency handoff mechanism which expedites network level handoff to provide real-time applications with an added advantage of seamless mobility.*

## 1. Introduction

The growing deployment of IEEE 802.11 [1] based (802.11a and 802.11b) wireless LANs in enterprise and public networks opens up opportunities for several novel applications and services which make use of the tetherless connectivity and mobility provided by these networks. Some of the examples of these services are; voice over WLANs, media streaming, last mile video distribution network, etc. The challenges involved in successful deployment of these services arise from the requirement of adequate bandwidth guarantee and effective roaming capability to support their real-time operation. All nodes in these wireless LANs use the same shared medium for communicating with each other. Collisions are inevitable when multiple nodes try to concurrently access this shared medium. These collisions are the primary reason for non-deterministic nature of these LANs. Further, the physical coverage of IEEE 802.11 based networks is limited because of the engineering constraints in the underlying radio technology. Collisions and physical coverage limitations act as hindrance in terms of adequate bandwidth and wide roaming capability.

QoS in wireless networks is a well recognized issue. The 802.11 specification extends rudimentary support in terms of a Point Coordination Function (PCF) towards QoS provisioning which is optional and vendors of WLAN equipments seldom choose to incorporate it in their devices. Further, this mechanism is not actively used by existing deployments. This co-ordination mechanism has been analyzed for performance mostly in simulated environments [2]. As a fallback approach, many schemes [3], [4], [5] have been pro-

*An earlier version of this work was published in Multimedia Computing and Networking (MMCN) conference in Jan 2002. This paper integrates this QoS scheme with a low latency mobility scheme that appeared in IEEE JSAC special issue on All IP Wireless Networks May 2004. This paper deals with both the issues with a fresh perspective of new networking technologies and standards such as 802.11e.

posed to support a priority mechanism over the available Distributed Co-ordination Function (DCF) of 802.11 wireless LANs. This is useful for applications which do not need hard QoS guarantees. Most of these priority mechanisms are based on Diffserv [6] in IP networks. The upcoming IEEE 802.11e [7] standard is aimed at improving QoS on 802.11a and 802.11b networks. It proposes a better way to handle time-sensitive traffic for multimedia applications. This is done by providing certain collision-free periods and accommodating time-scheduled and polled communications during this period. This mechanism is implemented in the form of so called Hybrid Co-ordination Function (HCF) and Enhanced Point Co-ordination Function (EPCF). However, these co-ordination functions are MAC layer capabilities and therein lies the problem of bridging the gap between link level co-ordination and application requirements. The problem becomes more prominent in the context of legacy applications which were not designed specifically for QoS oriented MAC protocols. To this effect, we describe a simple *software token* based bandwidth guarantee mechanism, called *Rether* [8], which is aimed at providing bandwidth guarantee to legacy applications by inter-operating with the network interface device drivers.

The physical coverage problem of wireless LANs is addressed by deploying multiple wireless LAN cells as access networks in an overlapped fashion where each cell is associated with an *access point*. These access points serve as layer-2 bridges between the high speed wired network and the wireless access network. As mobile nodes move in and out of these overlapped segments, they can associate with the corresponding access points according to beacon signal strengths. In 802.11b-based networks, the intelligence to measure signal strength and switch among network segments is built into the wireless LAN interface card, which exposes various status and control information to the software device driver. To enable cellular-like networking structure, wireless LAN adapters need to be configured to run in the *access point mode*, which is also known as the *infrastructure mode*. Mobile IP [9], an extension to the TCP/IP protocol suite, provides mobile nodes the ability to roam across wireless IP subnets without loss of network-layer connectivity. With Mobile IP, mobile nodes *do not need to reconfigure their IP addresses* while migrating across subnets. However, the mobility across subnets is not seamless from the purview of real-time applications. The latency to switch between different IP subnets is of the order of seconds in infrastructure mode wireless LANs. This latency appears as a period of loss of connectivity which is unacceptable for real-time applications. We describe a *low-latency* handoff extension to Rether, based on link handoff detection and network handoff expediation [10], resulting in a comprehensive solution towards *supporting bandwidth guarantee and mobility for real-time applications on wireless LANs*. We further describe how Rether can be adapted to emerging QoS oriented specifications of wireless LANs, namely, 802.11e.

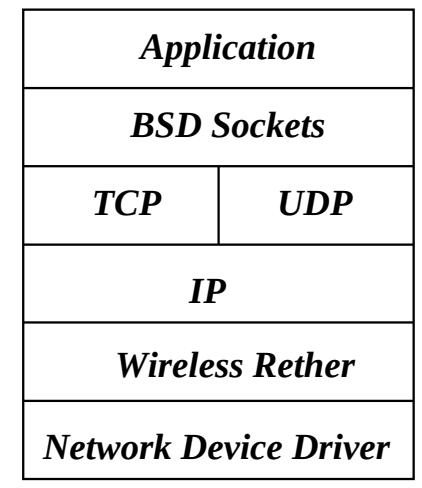

Fig. 1. Rether is implemented as a layer between the link-layer and the IP protocol layer. At this position, it can exercise QoS control over all outgoing traffic.

## 2. Design and System Architecture of Rether

Rether is a *software token* based protocol where a token is repeatedly circulated among the nodes of a wireless LAN. The token holder is given a contention free access to the channel to transmit data. This way the wireless network is collision-free by construction. The amount of data a node can transmit upon reception of the token depends on the bandwidth reservation of that node.

### 2.1. Hardware vs Software Implementation

Rether protocol could have been either implemented on the wireless LAN card or as a software layer above the network hardware's device drivers. The advantage of the hardware approach comes in the form of an extensive control over the token passing between nodes. Further, a hardware based implementation can effectively leverage the MAC channel co-ordination facility, such as, PCF or EPCF. The disadvantages of the hardware approach are that the solution becomes tied to individual vendor's hardware, the cost of hardware implementation is significantly higher, and there is less flexibility in configuring QoS policies. On the other hand, a software-based approach eliminates all these disadvantages. However, the software based approach faces a drawback in the form of additional token passing overhead. However, with careful protocol and software design, the token passing overhead can be kept to a minimum. This lead to a decision to adopt the software approach, whose structure is shown in Figure 1.

Another advantage of software based approach is its adaptability to standards like 802.11e. IEEE 802.11e defines two entities, namely, Station Management Entity (SME) and MAC Layer Management Entity (MLME), which provide a coordination between MAC layer and the other layers in network protocol stack for QoS provisioning. MLME is responsible for negotiating contention free channel access with a central channel co-ordinator called the *Hybrid Co-ordinator*. The software approach of Rether, sans the token passing part, can be easily adapted as SME for QoS provisioning in a setup where 802.11e based hardware is used.

### 2.2. Peer-to-peer vs. Centralized Token Passing

A central design decision in Rether was how the token is passed from one node to the next. One alternative is to maintain a logical ring among the wireless nodes and to implement a *peer-to-peer token passing* protocol. In the context of wireless LAN, since mobile nodes can move out of each other's transmission range, direct token passing between wireless nodes is infeasible. However, since all nodes are assumed to be reachable from the access point, the access point is in a better position to relay the token. As a result, we chose to implement a *centralized token-passing protocol* in which a central server, called *Wireless Rether Server* (WRS) is placed right between the access point and the wired network and is responsible for granting the token to wireless LAN nodes, called *Wireless Rether Clients* (WRC). This centralized architecture, has several advantages. First, it is the WRS and not the token that maintains the QoS-related state. Therefore loss of the token is not fatal to the proper functioning of the Rether protocol. Secondly, because the WRS can intercept all traffic entering and leaving the wireless LAN, it can snoop on the wireless channel to determine the end of a packet that a WRC sends and reduces the token passing overhead by eliminating the need of ACKs from WRCs.

Ideally one can combine the WRS and the access point into one device. This would have enabled us to integrate WRS functionality with the 802.11 channel coordinator resulting in a cleaner hardware setup. However, this integration is not currently possible because commercial off-the-shelf access points do not expose any programming interface to add third-party code. In retrospect, this limitation is a blessing in disguise, because the separate-WRS architecture requires the resulting Rether design to be independent of and thus able to inter-operate with 802.11 access points from multiple vendors.

### 2.3. Work-Conserving vs. Non-Work-Conserving

In Rether architecture, within a token passing cycle, the token first visits network nodes with bandwidth reservation, called real-time (RT) nodes, and after all the RT nodes have been visited the token visits the other network nodes that potentially have best-effort traffic to send, called non-real-time (NRT) nodes. Note that *every* wireless LAN node is an NRT node. If the token cannot visit all NRT nodes within a cycle, it continues to visit the NRT nodes in the next cycle starting from where it leaves off in the current cycle. Rether supports a non-work-conserving service model because even when there is no NRT traffic, the token is still passed among NRT nodes until the current cycle ends. Till the end of the cycle no RT node transmits RT data. That is, RT nodes can never send data at a higher rate than their reservation even when other RT or NRT nodes have less data to send. A non-work-conserving model reduces the extent of data burst and thus decreases the delay jitters that applications experience. In addition, this model fits well with the support for NRT or best-effort traffic.

### 2.4. Bandwidth Reservation, Packet Queuing and Scheduling

An important goal in developing Rether protocol was to support QoS for legacy and third party applications without modifying them. Since such applications cannot explicitly communicate their bandwidth requirements, Rether employs an indirect mechanism. Specifically, a *quintuple specification* of {`SrcAddress/Mask`, `DestAddress/Mask`, `SrcPortRange`, `DestPortRange`, `Bandwidth`} is mapped to reservation required by the legacy application in Rether policy tables. Except the bandwidth specification all other fields can be wildcards. If a matching specification is found in policy table for any new packet stream, a reservation request on behalf of the stream is sent to the WRS which then performs admission control. If the request is admitted, corresponding stream's packets are queued in a special real-time (RT) packet queue and dispatched by the network scheduler according to the QoS specifications. If the request is rejected, the stream is treated as a best effort.

Furthermore, Rether's QoS mechanism is designed to work seamlessly with the Diffserv [6] mechanism in the wide area networks. WRS interfaces Rether with wired Diffserv mechanism by transparently mapping upstream packets (destined to wired network) to one of the Diffserv classes and mapping downstream packets (destined to wireless network) to Rether's flow-specific reservation.

WRC classifies each outgoing packets as either *real-time* (with flow reservation), *urgent* (such as ICMP and ARP), *control* (Rether protocol) or *best-effort*. Upon token visit, WRC employs a *hybrid* packet scheduler in which the two limiting parameters are the reserved time share and the data share per cycle. The hybrid scheduler dispatches a flow's packets till either the flow's time share expires or its reserved share of data is transmitted. This scheme optimizes bandwidth usage for conservative reservations while dealing gracefully with bursty channel conditions. Additionally, WRC handles bidirectional nature of TCP connections by transparently setting up reverse channel reservation for TCP ACKs for the amount of 10% of forward data channel reservation.

## 3. Interoperability With 802.11e

IEEE 802.11e standard facilitates parameterized QoS, provided by means of an Enhanced Point Co-ordination Function (EPCF). In EPCF, a central channel coordinator, called Hybrid Coordinator (HC), allocates *transmission opportunities* (TxOP) to certain nodes during the so called *contention free periods*. The TxOP notification is done by means of poll frames sent by the HC. To provide TxOPs with appropriate duration at appropriate time, the HC needs to obtain the pertinent information from the individual nodes from time to time. For this purpose, the HC initiates the so-called controlled contention periods during which nodes send their resource requests to the HC without contending with other nodes which are sending only data traffic.

However, this QoS mechanism is at MAC layer and applications cannot directly utilize it without some interfacing mechanism which translates application requirements into appropriate transmission opportunities. More specifically, applications need a bandwidth reservation and scheduling mechanism that can utilize the MAC level TxOP functionality. Rether can readily leverage on 802.11e QoS functionality as the TxOP poll frames are exactly analogous to the software tokens in Rether. The bandwidth reservation and scheduling mechanisms can be integrated with EPCF without significant modifications. Thus, Rether can be used to bridge the gap between the MAC layer and applications. Further, Rether can bring in added QoS benefits because of the inherent priority support for non application specific urgent traffic, such as, ICMP, ARP, etc. The built-in support for TCP in terms of transparent reservation for ACK traffic provides additional QoS support.

## 4. Mobility Support

Rether has been designed ground up to seamlessly support higher level mobility protocols such as mobile IP. Support for host mobility with Rether has three distinct stages. First stage is a link-level handoff that occurs when a mobile node moves from one subnet to another and its network card detects the new access point in the vicinity. Second stage is a Rether-level handoff, in which the mobile node *registers* with the new WRS that announces its presence using periodic *beacon* messages. Final stage involves the regular mobile IP handoff to maintain IP-level session continuity across IP subnets.

Mobile IP handoff involves three entities: *home agent*, *foreign agent*, and the mobile node. Home agent and foreign agent are commonly referred as the mobile agents. These mobile agents periodically broadcast mobile IP advertisements on the wireless LANs. A mobile node, that moves from one IP subnet to another (foreign) subnet, intercepts these advertisements and sends a registration request to the newly discovered foreign agent. After due authentication an IP-over-IP tunnel is established between the home agent and the foreign agent. From this point onwards, the home agent intercepts and forwards all subsequent packets for the mobile node via the foreign agent. Similarly, all transmissions from the mobile node are forwarded upstream via the foreign agent to the home agent and finally to the destination.

During the mobile IP handoff the mobile node is essentially cut off from the network and transport layer connections appear to be stalled. This stalling is detrimental for the performance of real-time applications. Thus, it is imperative that the latency of handoff be minimized for smooth operation of real-time applications.

### *4.1. Handoff Latency Optimization*

It is feasible to reduce the mobile IP handoff latency by maintaining a list of potential foreign agents apriori and by anticipating the handoff based on the relative signal strength of the access points. Whenever a mobile node is about to migrate to a new subnet, it can contact the foreign agent serving the new subnet and carry out the handoff. This solution, though very simple and elegant, cannot work in the context of infrastructure mode of wireless LANs. The technical issues involved in this approach are that of maintaining a list of potential foreign agents apriori and anticipating the handoff. Unfortunately, existing 802.11 network interfaces do not support this mechanism. The link level handoff by 802.11 network interfaces is unilateral depending on the firmware logic and no notification is given to the software running on the mobile node about the impending link handoffs. Further, no information about signal strengths of access points from the adjoining cells is exposed to the higher protocol layers. This makes it impossible to anticipate and brace for impending handoffs in infrastructure setup.

Several schemes have been proposed to deal with handoff latency issues. Srinivasan Seshan et al. [11] propose a scheme which uses IP multicast and buffering to eliminate data loss and reduce the handoff delay. However, this scheme is based on *anticipating* a handoff using signal strength between the base stations and the mobile nodes. Also the handoffs are assumed to be initiated by mobile nodes. Though this scheme gives a very low handoff latency, it is not directly applicable in mobile IP deployments. Another approach is to use hierarchical mobility management as proposed by E. Gustafsson et al. [12] where a foreign administrative domain is covered by several foreign agents organized hierarchically. The local micro mobility of mobile nodes is masked from home network exposing only a top level gateway foreign agent to the home network. This approach improves the handoff performance by avoiding tunnel setup over the Internet whenever local handoffs occur. This scheme is complementary to the proposed scheme since it optimizes the tunnel setup time as opposed to link handoff detection and network handoff expediation.

E. Shim and R. Gitlin [13] propose a fast handoff mechanism, termed *NeighborCasting*, based on utilizing and sometimes even wasting the wired bandwidth in order to minimize the number of message exchanges between the wireless interfaces. This proposal requires modifications to Mobile IP to support the discovery of neighboring mobile agents which is used to create a neighborhood map for each mobile agent. Further, proposal assumes the availability of a notification mechanism from link layer to the Mobile IP layer about the impending handoff. In stable state, the neighborhood map is used to forward the data to all the neighboring mobile agents whenever a notification from the link layer is received. Thus a mobile node can brace for handoff and incur no data loss with some co-operation from neighboring mobile agents.

Almost all of the proposed solutions for reducing handoff latency invariably assume that mobile nodes can *anticipate* link-layer handoffs and maintain connectivity with the old as well as the new mobile agents during the handoff. This premise is not valid in the context of infrastructure mode wireless LANs using mobile IP as mobility management software.

Thus, in infrastructure setup with mobile IP, the only way to reduce the overall handoff latency is by *expediting* each of the following four stages in post link-level handoff processing: (1) time between link-layer handoff and its detection by the mobile IP software, (2) time from link-layer handoff detection to the reception of first foreign agent advertisement, (3) time for a mobile node to register with the new foreign agent detecting the first advertisement, and (4) time between request sent to and response returned from the new foreign agent.

### *4.2. Link-Layer Handoff Detection*

Although 802.11 network interface cards do not provide any mechanism to notify the software when link-layer handoff takes place, there is a hardware control functionality that allows software to probe for the identity of the access point with which a 802.11 card is currently associated. The wireless LAN MAC address of the access point is the identity reported by the 802.11 card. With this mechanism, software can detect change in the access point (and thus the occurrence of a link-layer handoff) by comparing results from consecutive probes. A mismatch from such comparison can then be used to trigger a mobile IP handoff.

These software probes themselves need to be triggered by certain events. The probing frequency determines how fast the occurrence of a link-layer handoff can be detected. Our empirical observations show that each link-layer handoff takes a few milliseconds. It does not make sense to probe for link-layer handoff at a higher frequency than once per several milliseconds. We therefore decided to piggyback this software probe with the system timer service routine, which is invoked once every 10 milliseconds. The cost of each probe takes only a few CPU clock cycles and does not add any noticeable performance overhead to the timer processing routine.

While the 802.11 cards support access point ID probing, the value that is returned in response to a probe is not always reliable. Sometimes the returned value is obviously ill-formed and sometimes it appears perfectly proper but turns out to be incorrect. Those proper-looking returned values lead to false positives and thus trigger two successive redundant mobile IP handoffs when in fact none should be triggered. Our experiences show that the number of false positives encountered increases with the probe frequency.

After analyzing the trade-off between the probe frequency and the reliability of return value, it was observed that one can maintain a high probe frequency without compromising the reliability of probed value by keeping a history of returned values from consecutive probes. More specifically, when two or three consecutive probes return the same value, it is very unlikely that the returned value is a false positive. We chose to implement such a history based probing mechanism.

### 4.3. *Foreign Agent Discovery*

Mobile IP specifies two mechanisms for mobile agent discovery. In the first mechanism, called *Mobile Agent Advertisement*, mobile agents periodically broadcast advertisement messages on the associated subnet to announce their presence to new mobile nodes. Since it is inadvisable to frequently broadcast advertisements over the entire subnet, the mobile IP specification limits the advertisements to one every second. The latency of mobile agent discovery based on advertisement mechanism is dependent on the frequency of the advertisements. Therefore, the latency of handoffs based on mobile agent advertisements is on the order of seconds, which is clearly not acceptable for real-time applications. Alternatively, mobile nodes, can send *Agent Solicitation* requests. However, the mobile IP standard requires mobile nodes to carry out solicitation only in the absence of agent advertisements and when the agent address could not be determined through any appropriate link-layer mechanism. The absence of an agent advertisement can be determined only when the lifetime of all the previously received advertisements has expired. Given that the lifetime of an advertisement is three times that of the advertisement interval, i.e. at least 3 seconds, it would take 1.5 seconds on the average to discover mobile agents after the link-layer handoff. Thus, with *Agent Solicitation*, the network-layer handoff time is still on the order of seconds.

Since mobile IP permits the use of any other *appropriate link-layer mechanism* to discover new mobile agent, we propose a *caching and replaying* approach to mobile agent solicitation. In this scheme, each subnet is augmented with a host on the wired segment acting as a *caching agent* which keeps the most recent mobile agent advertisement cached in its memory. Whenever a mobile node detects a link-layer handoff, it can send a solicitation packet to request for mobile agent advertisement. The caching agent, upon receiving a solicitation request responds back with the cached advertisement. This mobile agent solicitation scheme can help reduce the mobile agent advertisement frequency without increasing the handoff latency. Further, this scheme is independent of the mobile IP specification since the solicitation is for *replay* only and not for the advertisements.

### 4.4. *Integration with Rether*

At first glance, the caching and replaying scheme may seem an unnecessary extension when mobile IP itself provides a solicitation mechanism. However, this scheme proves to be beneficial when the solicitation and replay messages can be piggybacked with some other messages, such as, Rether registration and reply messages. the agent discovery can be performed in an overlapped manner with other message exchanges rather than in sequential manner. Advertisement soliciting, caching and replaying can all be done in a way completely transparent to mobile IP. The net result is a transparent and independent mobile Agent discovery process. The *caching agent* need not be an independent host on the wired network. The *caching and replay* module may be run on the WRS in Rether setup. We integrated the proposed scheme with Rether in order to provide QoS with roaming support. In this integrated setup, the WRS acts as the caching agent. Whenever a mobile node (WRC) migrates to a new wireless subnet, it registers with the WRS responsible for the new subnet by sending a *rether registration* request. The solicitation for the cached Mobile IP advertisement is piggybacked with the registration request. The WRS then piggybacks the cached advertisement with the *rether registration reply*. Also, the QoS mechanism treats the Mobile IP *registration* and *reply* messages as *urgent* messages in order to expedite the Mobile IP processing. After registering with the new WRS, the WRC reestablishes bandwidth reservations for each of its active application. The system architecture of this integrated system is shown in Figure 2.

## 5. Performance Measurements

We evaluated the Rether protocol by testing a fully functional prototype implemented on Linux platform. The implementation was tailored for 802.11b wireless LANs. For 802.11b networks, the peak bandwidth possible is 11 Mbps. But the normal link bandwidth that can be observed is around 6.5 Mbps when a single sender is transmitting 1500 byte packets in the absence of collisions. When there are multiple senders the collisions result in packet loss and hence reduction in the overall throughput.

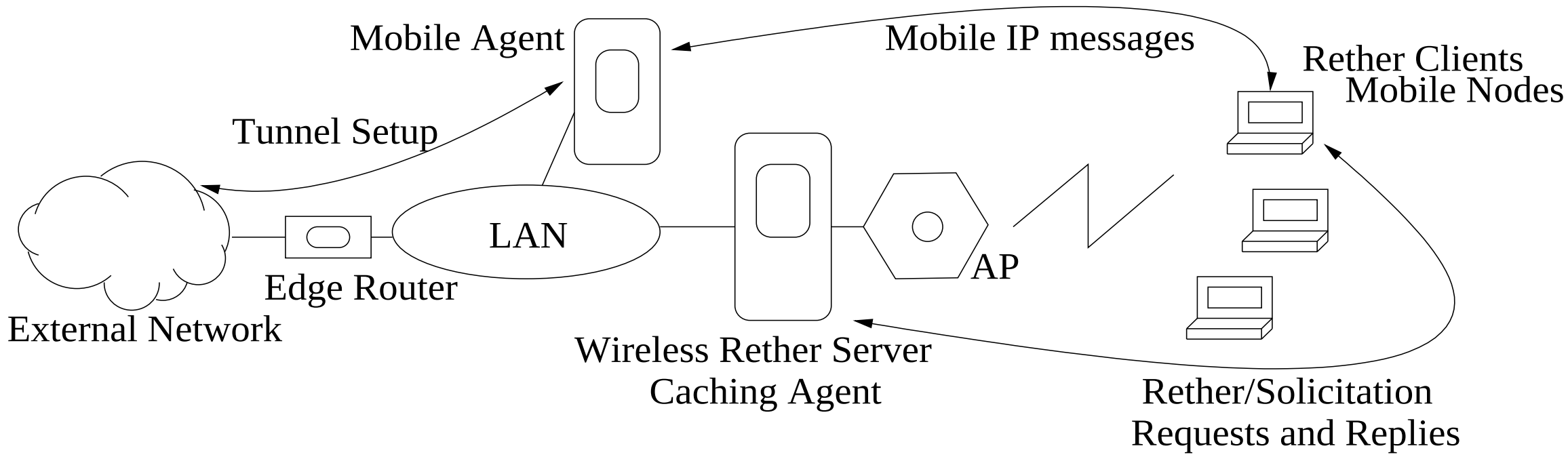

Fig. 2. *Integration of Fast Handoff scheme with Rether. The WRS acts as the caching agent. The arrows indicate the messages exchanged between different nodes. The initial solicitation requests are piggybacked over Rether messages. Subsequent messages are treated as urgent messages by Rether.*

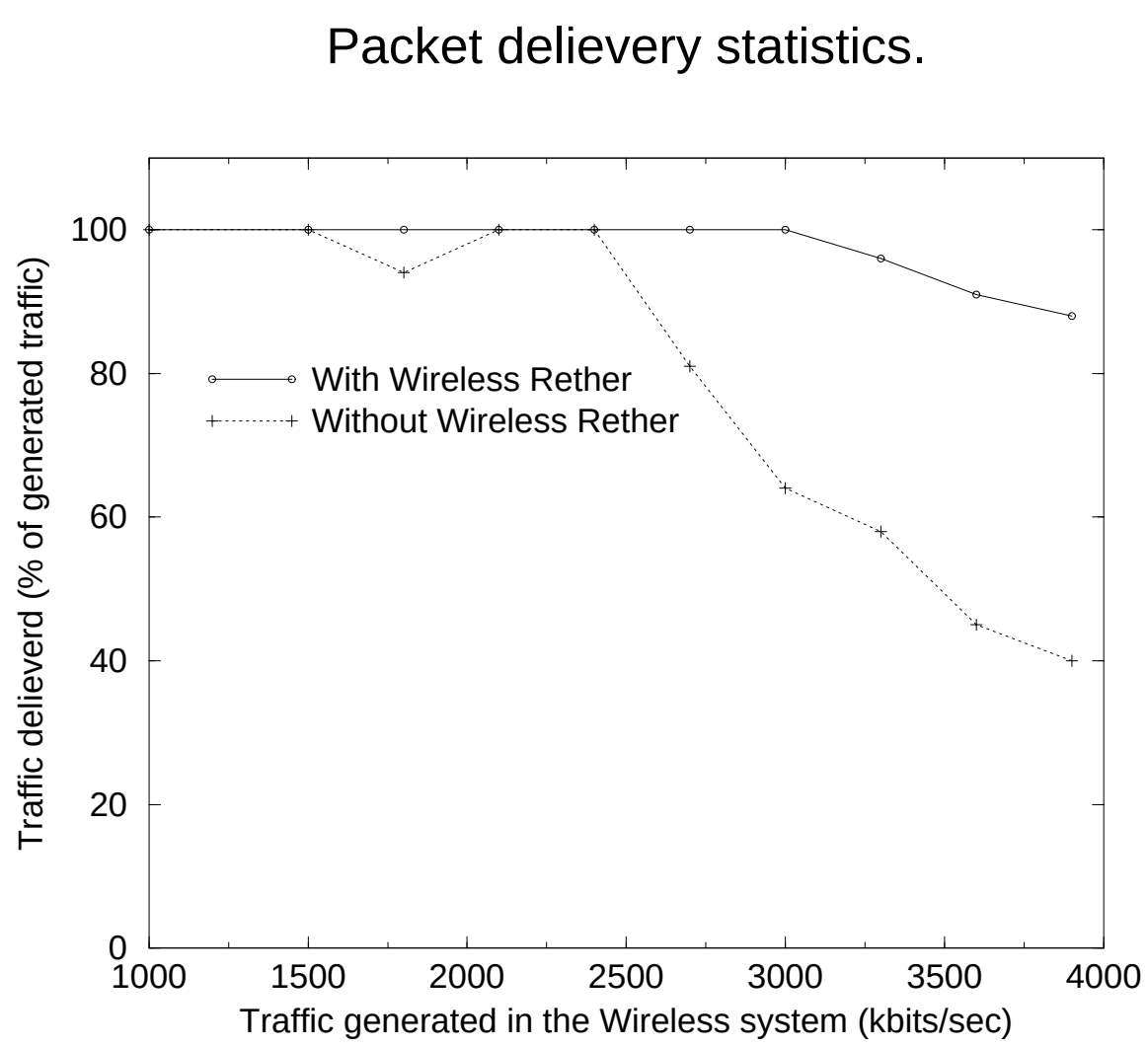

Fig. 3. Packet loss with three senders transmitting. Rether reduces packet loss from 60% to a mere 10%.

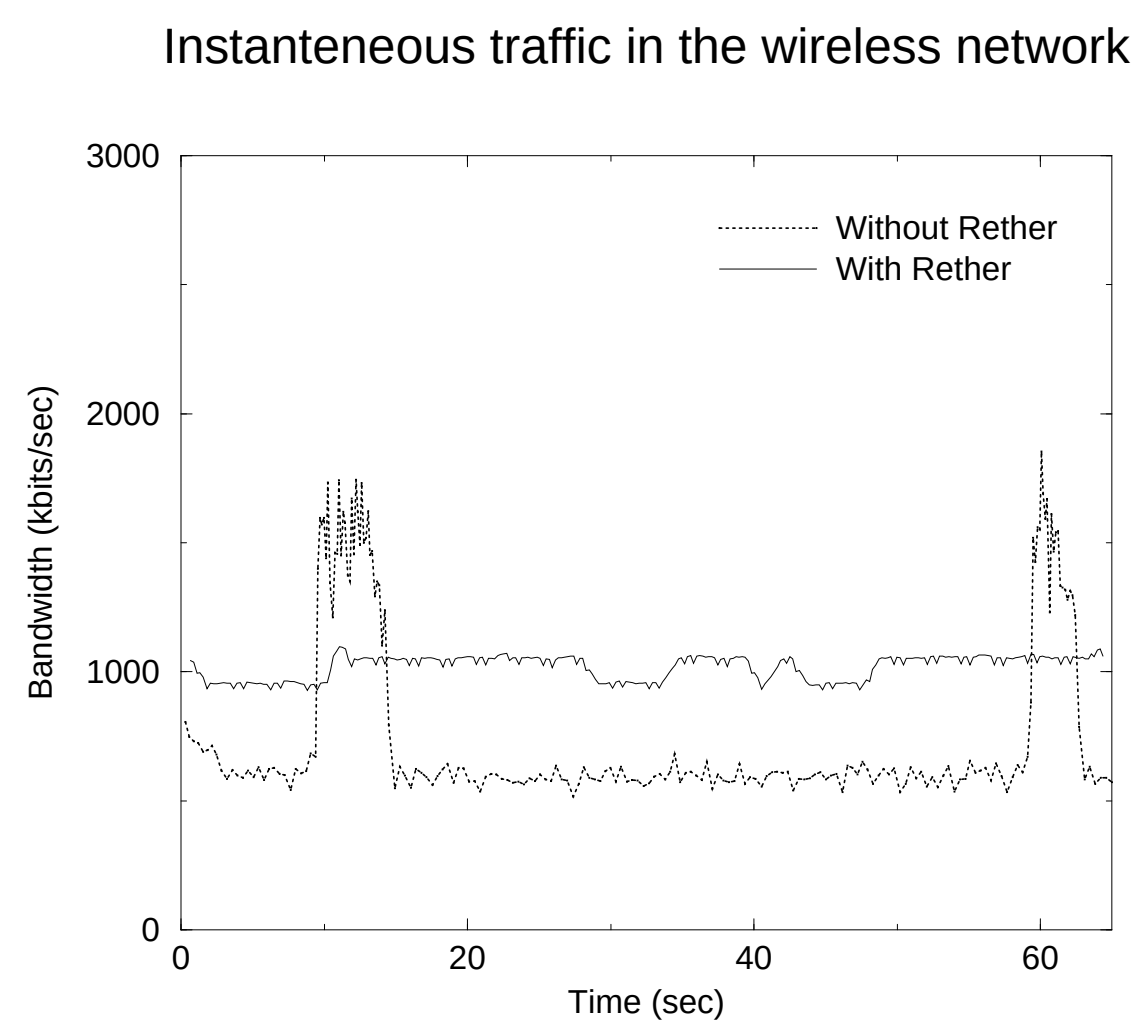

Fig. 4. Rether reduces the the bursty nature of transmissions and increases the throughput.

Our experimental setup consisted of three hosts transmitting simultaneously - two of them upstream and one downstream. Figure 3 shows a comparison of packet loss observed in the presence and absence of Rether. Without Wireless Rether, we can observe packet losses even at traffic load of 2.5 Mbps, beyond which packet losses increase drastically to 60% when the load is around 4 Mbps. In contrast, with Rether the packet drops stay within 10% of the total network load. The primary reason for packet losses in Rether was that the access point was unable to handle heavy loads, and thus resulted in dropped packets.

Another advantage of Rether is that it reduces the bursty nature of the traffic in the network. Figure 4 shows the throughputs of a wireless network in the presence and absence of Rether. In both cases three senders attempt to transmit data at a rate of 1.1 Mbps. Without Rether the traffic is bursty and the actual throughput of each connection is lower than the transmission rate of 1.1 Mbps. In the presence of Rether, the required throughput is achieved and the burstiness is reduced as well.

The time to detect the link-layer handoff is determined by the probe interval. Since our detection scheme requires *at least* two probes to confirm any handoff, each detection takes at least two invocations of the probe routine. The handoff detection time, measured from the beginning of the last handoff, is within a relatively short range of 50 milliseconds to 70 milliseconds. Immediately after a link-layer handoff detection, a mobile node can send out a solicitation for the new mobile agent. The response to this solicitation is received after a processing and message round-trip delay of 4 milliseconds. This solicitation response triggers the Mobile IP registration, which is received by the mobile agent within the next two milliseconds. Further authentication and tunnel setup was observed to take around 36 milliseconds in LAN setup. Therefore, the amount of time required for a Mobile IP handoff is about 100 milliseconds in the average case (60 msec + 4 msec + 36 msec).

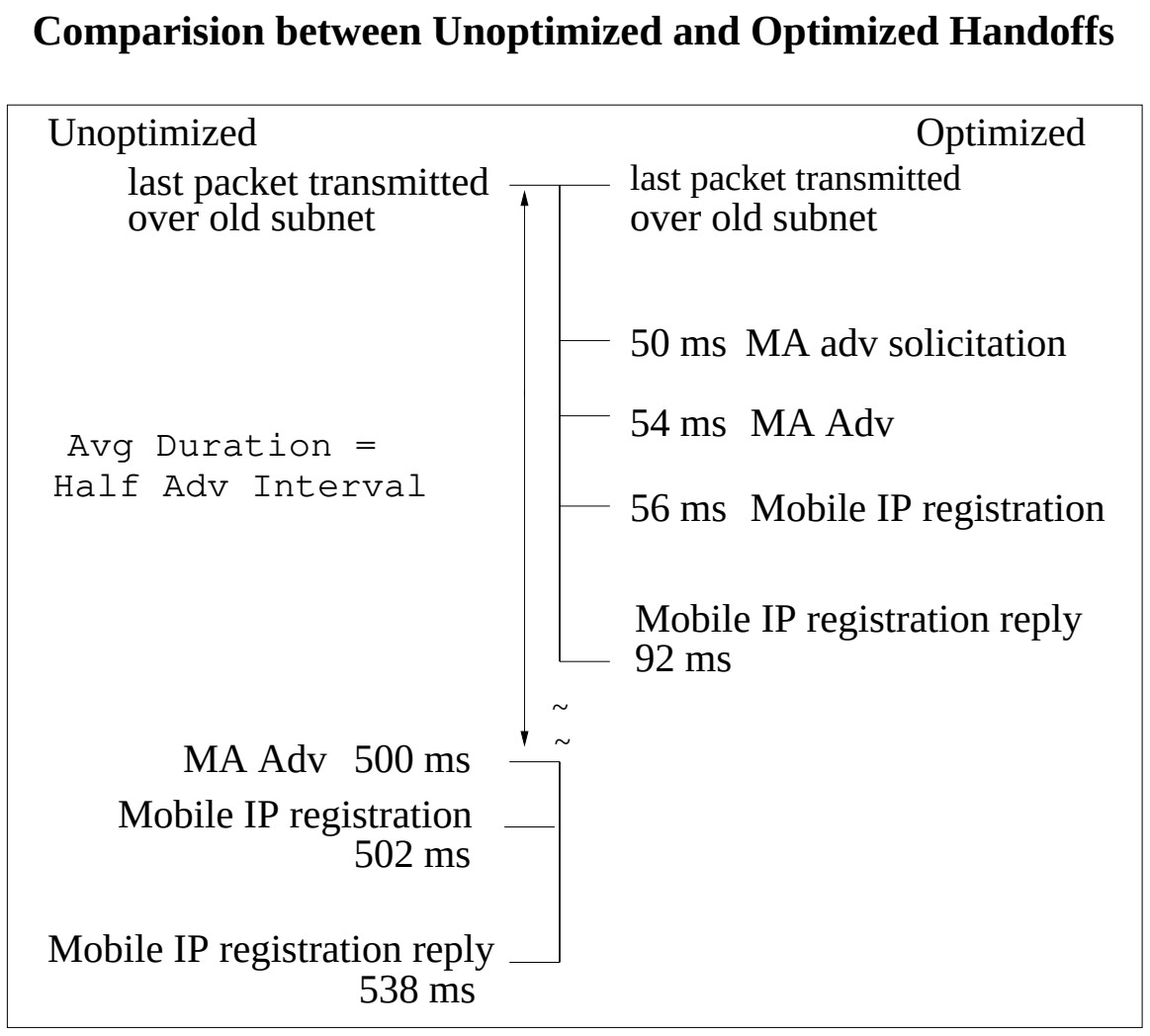

Fig. 5. *Gant chart for the Mobile IP handoff with and without detection optimization. In the optimized case, the handoff time is around 100 milliseconds, whereas, in unoptimized case the average handoff time is around 550 milliseconds and can extend up to several seconds.*

In contrast to this optimized scheme, the regular Mobile IP handoff implementation initiates a network-layer handoff only upon reception of a mobile agent advertisement. When the advertisement period is allowed minimum of 1 second, the average delay between the completion of a link-layer handoff and reception of the next mobile agent advertisement is around 500 milliseconds. Subsequent Mobile IP processing may take around 35 to 50 milliseconds. Thus, the *average* Mobile IP handoff latency measured from the reception of the last packet on previous subnet is around 550 milliseconds. The proposed optimization reduces the average handoff latency by more than 450 milliseconds, which is significant for time-critical applications. The deviation in the handoff latency for the unoptimized case is around 500 milliseconds, whereas, for the optimized case it is only 10 milliseconds. Thus the upper bound on the Mobile IP handoff latency is of the order of seconds in the unoptimized case and is still below 110 milliseconds in the optimized case. Figure 5 shows the timing diagram that contrasts the optimized and unoptimized implementations for Mobile IP handoff.

## 6. Conclusion

In this paper we described the design of Rether protocol that provides QoS guarantees on wireless LANs. We also described the integration of a low-latency handoff mechanism for mobile IP with Rether. Rether adopts a centralized token passing architecture, which can be adapted to forthcoming 802.11e specifications. Rether ensures that the data transmission on wireless LANs is collision free in order to provide bandwidth guarantee. It uses port-based reservation mechanism to support legacy applications. It transparently interoperates with higher layer protocols like mobile IP to support roaming capabilities while incurring a very low latency during handoff process. We also showed the performance measurements which validate the effectiveness of Rether to provide QoS on wireless LANs.